\documentclass[trackchanges,twocolumn]{aastex701}

\begin{document}

\title{A Sample of 459 Galaxy Groups Identified from the FAST All Sky H\,{\sc i} Survey DR2}

\author[gname=Zhongsheng, sname='Yuan']{Z. S. Yuan}
\affiliation{National Astronomical Observatories, Chinese Academy of Sciences, 20A Datun Road, Chaoyang District, Beijing 100101, China}
\affiliation{CAS Key Laboratory of FAST, NAOC, Chinese Academy of Sciences, Beijing 100101, China}
\affiliation{School of Astronomy, University of Chinese Academy of Sciences, Beijing 100049, China}
\email[show]{zsyuan@nao.cas.cn}  

\author[gname=Zhonglue, sname='Wen']{Z. L. Wen}
\affiliation{National Astronomical Observatories, Chinese Academy of Sciences, 20A Datun Road, Chaoyang District, Beijing 100101, China}
\affiliation{CAS Key Laboratory of FAST, NAOC, Chinese Academy of Sciences, Beijing 100101, China}
\affiliation{School of Astronomy, University of Chinese Academy of Sciences, Beijing 100049, China}
\email[]{zhonglue@nao.cas.cn}

\begin{abstract}

Groups of H\,{\sc i}-rich galaxies are key laboratories for
understanding galaxy evolution and environmental effects, constructing
samples of such groups is fundamental for relevant studies. Using the
FAST All Sky H\,{\sc i} Survey (FASHI) Data Release 2 (DR2), currently
the largest and deepest H\,{\sc i} catalog available, we identify
galaxy groups directly from H\,{\sc i} signals, without relying on
optical pre-selection. A group-finding algorithm with a density
contrast criterion and a minimum membership of five yields a catalog
of 459 galaxy groups containing 3015 H\,{\sc i} sources. Using
archival spectroscopic data, we identify optical counterparts and
derive total stellar masses of member galaxies as a proxy for the
group total mass. The group redshift distribution peaks at \(z \sim
0.02\), and the total H\,{\sc i} mass peaks at \(\sim 10^{10}
M_{\odot}\). We find a significant positive correlation between total
H\,{\sc i} mass and total stellar mass, consistent with the increase
in H\,{\sc i} mass with halo mass reported in previous stacking
studies. The ratio of total H\,{\sc i} mass to total stellar mass
decreases significantly with increasing total stellar mass, indicating
gradual H\,{\sc i} depletion in groups. Our sample provides rich
targets for future studies of galaxy evolution and large-scale
structure in the local Universe.

\end{abstract}



\section{Introduction} 

The gas content of galaxies is regulated by both internal feedback and
external environmental processes \citep[e.g.,][]{bg06,no17}. Among the
different environments, galaxy groups occupy a special role: they are
less massive and less dynamically evolved than clusters, yet they
contain a significant fraction of the galaxy population and are often
the first place where environmental effects become visible. In these
systems, the neutral atomic hydrogen (H\,{\sc i}) reservoir is
particularly fragile. It can be removed by ram-pressure stripping,
truncated by tidal interactions, or slowly consumed after the gas
supply is cut off \citep[e.g.,][]{gg72,mkl+96,bnm00,kvv04}. Observing
how the H\,{\sc i} content of groups changes with group mass is
therefore a direct way to test the onset and efficiency of these
processes.

H\,{\sc i} investigations on individual galaxy groups have provided
detailed views of gas removal in specific systems. Interferometric
surveys of groups such as those by \citet{fsw09}, \citet{kfb+09},
\citet{pbs+11}, and \citet{woc+17} have generally found flat low-mass
slopes (\(\alpha \approx -1\)) in the group H\,{\sc i} mass function,
in contrast to the steeper global value. A few groups, however,
exhibit steeper slopes \citep{shg+09,dab+11}, indicating that
group-to-group variations can be significant. These studies highlight
the diversity of H\,{\sc i} content in groups, but their small sample
sizes and heterogeneous selection make it difficult to draw general
conclusions.

Most previous measurements of the H\,{\sc i} content of a sample of
galaxy groups have started from optically defined group catalogs and
then either matched or stacked the H\,{\sc i} data. For example,
\citet{gjh+20} stacked H\,{\sc i} spectra from the Arecibo Legacy Fast
ALFA (ALFALFA) survey for groups based on the Sloan Digital Sky Survey
(SDSS) and found that the total H\,{\sc i} mass increases with halo
mass, with an additional dependence on halo richness. \citet{ddm+23}
combined groups from the Galaxy And Mass Assembly (GAMA) survey with
ALFALFA data and derived the H\,{\sc i}-to-halo mass relation, showing
that the group H\,{\sc i} mass continues to rise up to halo masses of
\(10^{13.7}M_{\odot}\) while the H\,{\sc i} mass fraction
declines. \citet{jha+20} further showed that the H\,{\sc i} mass
function of group galaxies has a flatter low-mass slope than the
global H\,{\sc i} mass function. These studies have been highly
informative, but optical selection can bias the group membership,
especially for gas-rich galaxies with faint stellar counterparts.

Blind 21-cm surveys offer a complementary approach by selecting
galaxies directly from their H\,{\sc i} emission. The H\,{\sc i}
Parkes All-Sky Survey (HIPASS) and ALFALFA established the local
H\,{\sc i} mass function and the cosmic H\,{\sc i} density
\citep{zms+05,mpg+10,jhg+18}. More recently, the FAST All Sky H\,{\sc
  i} Survey (FASHI) has pushed both sensitivity and sky coverage much
further. Its second data release (DR2) contains 156,411 extragalactic
H\,{\sc i} sources over about \(19,500~\mathrm{deg}^2\), with a median
sensitivity of \(0.57~\mathrm{mJy\,beam^{-1}}\) at
\(6.4~\mathrm{km\,s^{-1}}\) resolution \citep{zzj+24,zzj+26}. Because
the catalog is built from a blind H\,{\sc i} search and has a
well-characterized completeness, it provides an opportunity to
identify galaxy groups directly from their H\,{\sc i} emission rather
than from an optically selected parent catalog.

In this paper, we build the first galaxy group sample selected
directly from a blind H\,{\sc i} survey without any optical
pre-selection. Using FASHI DR2, we apply a group-finding algorithm
based on a density contrast criterion and require at least five
H\,{\sc i} members. This yields 459 groups containing 3015 H\,{\sc i}
sources. We then identify optical counterparts with spectroscopic
redshifts from the Dark Energy Spectroscopic Instrument (DESI) Data
Release 1 (DR1), the SDSS DR18, the Two Micron All Sky Survey (2MASS),
and the NASA/IPAC Extragalactic Database \citep{ned}, and use
the total stellar mass of the confirmed member galaxies as a proxy for
the total mass of the group. With this sample, we examine how the
total H\,{\sc i} mass of a group varies with its total stellar mass,
and how the ratio of total H\,{\sc i} mass to total stellar mass
changes with group mass. The paper is organized as
follows. Section~\ref{sect2} describes the FASHI DR2 data, the group
identification procedure, and the identification of optical
counterparts. Section~\ref{sect3} presents the redshift and mass
distributions of the groups and the relations between total H\,{\sc i}
mass and total stellar mass. We summarize our results in
Section~\ref{sect4}. Throughout, we adopt a flat \(\Lambda\)CDM
cosmology with \(H_0 = 70~\mathrm{km\,s^{-1}\,Mpc^{-1}}\), \(\Omega_m
= 0.3\), and \(\Omega_\Lambda = 0.7\).

\section{Identification of H\,{\sc i} galaxy groups}
\label{sect2}
\subsection{The FAST All Sky H\,{\sc i} Survey}
\label{data}
The FASHI survey is a wide-field H\,{\sc i} survey carried out with
the Five-hundred-meter Aperture Spherical radio Telescope
\citep[FAST,][]{nlj+11}. Its primary goal is to map the entire sky
visible from FAST, covering declinations from $-14^\circ$ to
$+66^\circ$ and a total area of about 22,000\,deg$^2$, over the
frequency band 1.0--1.5\,GHz. Using the 19-beam receiver in drift-scan
mode, FASHI combines broad sky coverage with excellent
sensitivity. The first data release (DR1), based on observations taken
between August\,2020 and June\,2023, spans more than 7,600\,deg$^2$
and yields 41,741 extragalactic H\,{\sc i} sources within the
frequency window 1305.5--1419.5\,MHz, reaching redshifts up to
$z\approx0.09$ \citep{zzj+24}. DR1 achieves a median sensitivity of
$\sim$0.76\,mJy\,beam$^{-1}$ at a spectral resolution of
6.4\,km\,s$^{-1}$. By cross-matching with the Siena Galaxy Atlas
\citep[SGA,][]{mld+23} and the SDSS \citep{aaa+09}, about 40.7\% of
the DR1 sources are assigned spectroscopic redshifts, while another
26.3\% have photometric redshifts.

The second data release (DR2) substantially extends the survey in both
depth and area \citep{zzj+26}. It now covers roughly 19,500\,deg$^2$
and lists 156,411 extragalactic H\,{\sc i} sources, with a median
sensitivity improved to 0.57\,mJy\,beam$^{-1}$ while preserving the
6.4\,km\,s$^{-1}$ spectral resolution. DR2 not only fully includes the
DR1 footprint but also overlaps the ALFALFA region \citep{hgm+11} and
adds detections at lower velocities. Compared with ALFALFA, FASHI DR2
achieves higher sensitivity (0.57 \,mJy\,beam$^{-1}$ with a velocity
resolution of 6.4 \,km\,s$^{-1}$ vs.~1.8\,mJy\,beam$^{-1}$ with 10
\,km\,s$^{-1}$) and nearly double the source density \citep[$\sim$8.0
  vs.~4.8\,deg$^{-2}$,][]{ghk+05}. Importantly, the DR2 catalog has
undergone a rigorous completeness analysis that accounts for the
survey's non-uniform sensitivity and line-width dependence, providing
reliable completeness estimates for each source \citep{zzj+26}. The
catalog also exhibits a high purity, with a false-detection rate
estimated to be below 2\% based on multi-criteria visual inspection
and cross-matching with spectroscopic surveys. These characteristics
make FASHI DR2 the largest and deepest H\,{\sc i} catalog currently
available, offering an invaluable dataset for studying the neutral gas
content, galaxy evolution, and large-scale structure in the nearby
Universe.

\subsection{Group Identification Method}
\label{method}

\begin{figure*}
  \centering
      {\includegraphics[angle=0,width=0.6\textwidth]{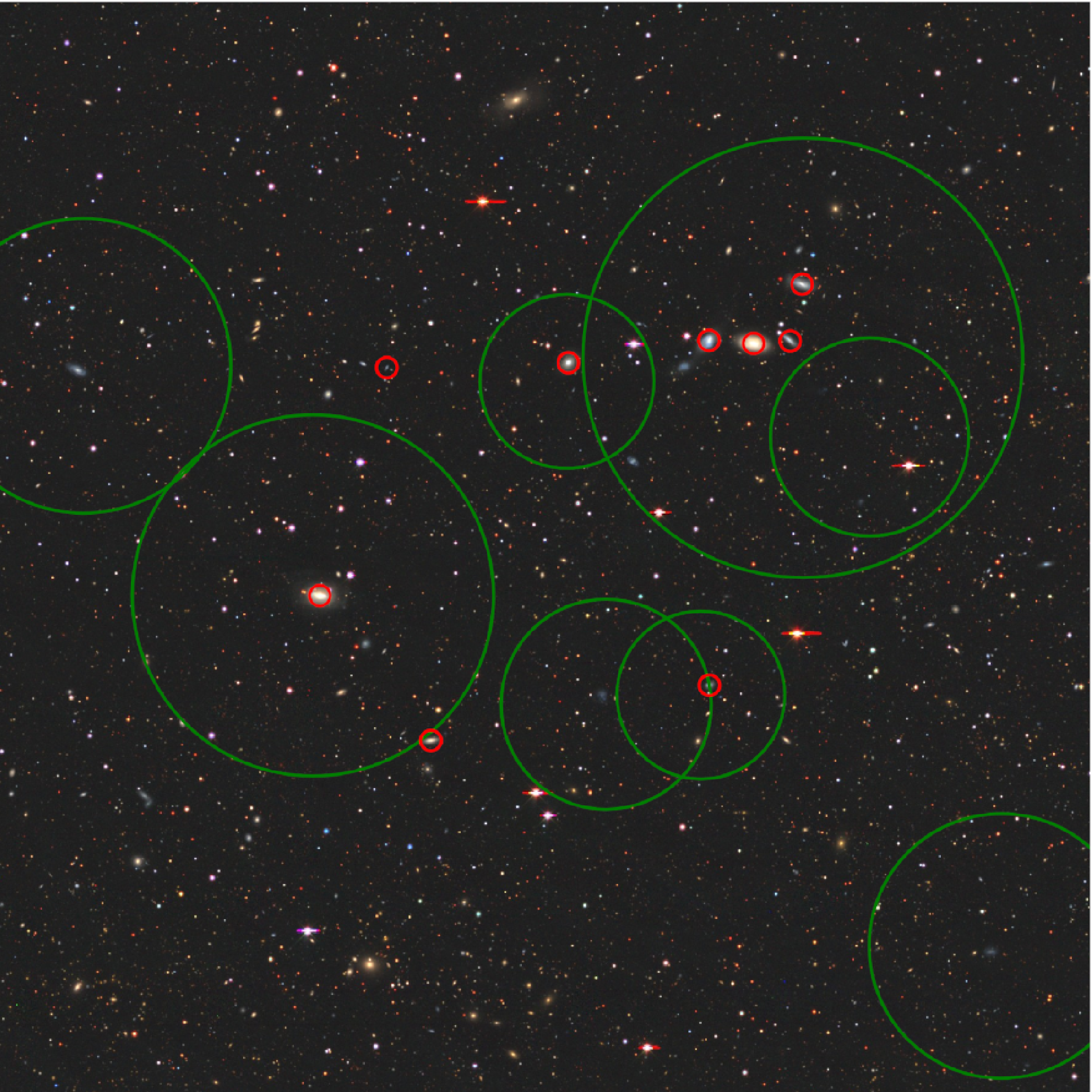}}
      \caption{Optical image of the galaxy group with ID = 1,
        retrieved from the DESI Legacy Survey DR11. Large circles
        indicate the apertures over which the flux of each H\,{\sc i}
        source is integrated \citep[see][for details]{zzj+26}, while
        small circles mark the optical spectroscopically confirmed
        member galaxies of this group (see section~\ref{counterpart}
        for details). The group is at a redshift of $z=0.0178$, and
        the side length of the image corresponds to an angular scale
        of 36.8 arcmin and a physical scale of 800 kpc.}
      \label{fig1}
\end{figure*}

We construct a catalog of galaxy groups based on the H\,{\sc i} sample
of FASHI DR2. The identification procedure consists of three steps,
designed to reliably extract physically associated galaxy groups from
the survey data.

For each galaxy as a potential center, we first search for companion
galaxies within a projected radius of $r = 300\,$kpc and a velocity
window of $\Delta v = 500\,$km\,s$^{-1}$. This choice of parameters is
motivated by the typical virial radius ($\sim260\,$kpc) and central
escape velocity ($\sim460\,$km\,s$^{-1}$) of a low-mass dwarf galaxy
group of $\sim10^{12}\,M_\odot$ \citep[e.g.,][]{slj+17,ywh26},
ensuring adequate coverage of potential physically associated members
in low-mass groups. To exclude spurious groups arising from random
projection effects, we impose a number density contrast criterion: the
number density within the 300\,kpc region must be at least three times
that within the radius of 1\,Mpc, i.e., $\rho_{300\rm kpc} \ge
3\rho_{1\rm Mpc}$. Additionally, we require a minimum membership of
$N_{\mathrm{mem}} \ge 5$ (including the center) to ensure statistical
robustness.

Because some candidate groups may share members, we merge candidate
groups that have common members into unified group candidates. For
each merged group, we define its center as the spatial density center
as follows: for each member, we compute the sum of angular distances
to its four nearest neighbors, and adopt the galaxy with the minimum
sum as the center. This definition is insensitive to outliers and
provides a good representation of the mass concentration of the group.

\begin{deluxetable*}{crrrrrrrccc}
\tabletypesize{\small}
\tablecaption{Parameters of galaxy groups with no less than five H\,{\sc i} members (the full sample is available at the CDS).}
\tablehead{
\colhead{ID} & \colhead{RA} & \colhead{DEC} & \colhead{$v$} & \colhead{$D$} & \colhead{$z$} & \colhead{log$_{10}M_{\rm HI,tot}$} & \colhead{$\langle C \rangle$} & \colhead{$N_{\rm HI}$} & \colhead{$N_{\rm s}$} & \colhead{log$_{10}M_{\rm *,tot}$} \\
 & \colhead{(J2000)} & \colhead{(J2000)} & \colhead{($\rm km~s^{-1}$)} & \colhead{(Mpc)} & & \colhead{($M_{\odot}$)} & & & & \colhead{($M_{\odot}$)} \\
(1) & \colhead{(2)} & \colhead{(3)} & \colhead{(4)} & \colhead{(5)} & \colhead{(6)} & \colhead{(7)} & \colhead{(8)} & \colhead{(9)} & \colhead{(10)} & \colhead{(11)}}
\startdata
 1  & 0.30522  & 13.10330  & 5250.2$\pm$1.4  & 72.1$\pm$5.4  & 0.0178 & 10.40$\pm$0.04  & 0.94 &  8  &  9  & 11.12 \\
 2  & 0.36546  & 31.49907  & 4752.7$\pm$1.4  & 60.6$\pm$4.5  & 0.0159 &  9.46$\pm$0.04  & 0.72 &  5  &  2  & 10.88 \\
 3  & 1.20084  & -1.56815  & 6997.5$\pm$0.8  & 96.8$\pm$4.8  & 0.0237 & 10.55$\pm$0.02  & 0.97 &  5  &  5  & 10.61 \\
 4  & 1.40655  &  8.60337  & 4961.0$\pm$2.5  & 60.7$\pm$4.6  & 0.0171 & 10.19$\pm$0.04  & 0.98 &  7  &  9  & 10.73 \\
 5  & 1.96208  & 15.75583  &  849.1$\pm$1.8  & 11.2$\pm$0.8  & 0.0031 &  9.52$\pm$0.03  & 1.00 & 12  &  7  & 10.42 \\
 6  & 3.50454  & 48.23558  & 4927.2$\pm$0.5  & 62.5$\pm$4.7  & 0.0167 &  9.93$\pm$0.04  & 0.81 &  5  & --  &  --   \\
 7  & 5.20125  & 22.06222  & 5656.7$\pm$2.4  & 78.5$\pm$5.9  & 0.0189 & 10.12$\pm$0.03  & 0.76 &  7  &  3  & 10.97 \\
 8  & 5.36856  & 22.49403  & 5661.1$\pm$3.6  & 78.5$\pm$5.9  & 0.0181 & 10.03$\pm$0.04  & 0.75 &  6  &  7  & 11.56 \\
 9  & 6.72866  & 49.12095  & 5145.3$\pm$0.7  & 77.5$\pm$5.8  & 0.0179 & 10.09$\pm$0.03  & 0.92 &  6  & --  &  --   \\
10  & 6.85971  & -1.66247  & 4043.4$\pm$1.6  & 50.1$\pm$3.8  & 0.0136 & 10.23$\pm$0.04  & 0.97 &  9  &  9  & 10.89 \\
 ... & ... & ... & ... & ... & ... & ... & ... & ... & ... & ... \\
 \enddata
\tablecomments{Columns: (1) Group ID. (2 - 5) Right ascension, declination, velocity and distance of the central H\,{\sc i} source. These parameters are directly taken from the FASHI catalog \citep{zzj+26}. (6) Group redshift, derived from the mean velocity of the H\,{\sc i} sources within the group. (7) Total H\,{\sc i} mass of the group, which is the sum of H\,{\sc i} masses of all member sources. (8) H\,{\sc i} mass-weighted mean completeness of the group. (9) Number of H\,{\sc i} sources in the group. (10) Number of optically confirmed member galaxies in the group. "--" indicates that the group lies outside the DESI DR1 footprint. (11) Total stellar mass of optically confirmed member galaxies.}
\end{deluxetable*}

To more accurately determine the boundaries of each group, we perform
three iterations of refinement for each merged group. In the first
iteration, starting from the central galaxy defined above, we collect
candidate members within a radius of $r = 1\,$Mpc and a velocity
window of $\Delta v = 500\,$km\,s$^{-1}$. We then apply the same
number density contrast criterion to determine the critical radius
$r_{\mathrm{max}}$, defined as the projected distance to the farthest
member that satisfies the density condition $\rho_{300\rm kpc} \ge
3\rho_{1\rm Mpc}$, and recompute the density center. We then compute
the dynamical mass $M_{\mathrm{dyn}}$ of the candidate group following
\citep[e.g.,][]{htb85}
\begin{equation}
  \label{dynmass}
  M_{\rm dyn}=\frac{32}{\pi G(N-3/2)}\sum^N_ir_i\Delta v_i^2,
\end{equation}
where $N$ is the number of member galaxies, $r_i$ and $\Delta v_i$ are
the distance and velocity of each galaxy relative to the central
galaxy, $G$ is the gravitational constant. Given that H\,{\sc i}
galaxy groups are generally dynamically young, the dynamical method
tends to overestimate their masses; we therefore adopt the assumption
that $M_{\mathrm{vir}} = \frac{1}{2}M_{\mathrm{dyn}}$
\citep[e.g.,][]{fbc17}. Then the central escape velocity
$v_{\mathrm{esc}}$ can be calculated through \citep[e.g.,][]{ywh26}
\begin{equation}
  v_{\rm esc}=\sqrt{\frac{2GM_{\rm vir}}{r_{\rm
        vir}}\cdot\frac{c_{\rm vir}}{{\rm ln}(1+c_{\rm vir})-c_{\rm
        vir}/(1+c_{\rm vir})}},
\end{equation}
where $c_{\rm vir}$ is the group concentration that is derived from
the mass-concentration relation of \citet{dsk+08}:
\begin{equation}
c_{\rm vir}=7.85\left(\frac{M_{\rm
    vir}}{2\times10^{12}h^{-1}
  M_{\odot}}\right)^{-0.081}\left(1+z\right)^{-0.71},
\end{equation}
where $h=H_0/100$ and $z$ means the group redshift. In the second
iteration, we update the search radius to $r=2\,r_{\mathrm{max}}$ and
the velocity window to $\Delta v =v_{\mathrm{esc}}$, and also the
density contrast selection $\rho_{r_{\mathrm{max}}} \ge
3\rho_{2r_{\mathrm{max}}}$, and obtain a new $r'_{\mathrm{max}}$ and
density center. The third iteration repeats the same procedure. These
three iterations allow the group boundary to adapt to the actual
density distribution and use dynamical information to constrain the
velocity range, effectively excluding unrelated galaxies in both
projection and velocity space. We retain groups with $N_{\mathrm{mem}}
\ge 5$ as the final group sample. Finally, we obtain 459 galaxy
groups, containing a total of 3\,015 H\,{\sc i}
sources. Figure~\ref{fig1} presents the image of the group with $\rm
ID=1$ as an example. Table~\ref{tab1} lists parameters of these 459
groups, while the parameters of the 3\,015 H\,{\sc i} members are
listed in the Appendix.

We assess the completeness of the total H\,{\sc i} mass of each group
using the per-source completeness $C_i$ provided in FASHI DR2. Since
the total H\,{\sc i} mass of each group is dominated by its massive,
high-SNR members with $C_i \simeq 1$, we compute the mass-weighted
mean completeness for each group as $\langle C\rangle = \sum_i M_{{\rm
    HI},i} C_i / \sum_i M_{{\rm HI},i}$. We find that 70.6\% of the
groups have $\langle C\rangle > 0.9$ and 84.1\% have $\langle C\rangle
> 0.8$, indicating that the total H\,{\sc i} masses of our groups are
largely complete.

\subsection{Identification of optical counterparts}
\label{counterpart}

\begin{figure*}
  \centering
      {\includegraphics[angle=0,width=0.8\textwidth]{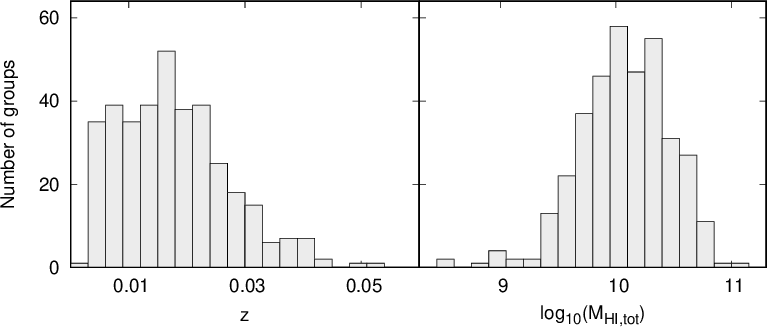}}
      \caption{Distributions of redshift (left panel) and total
        H\,{\sc i} mass (right panel) of the 459 identified galaxy
        groups.}
      \label{fig2}
\end{figure*}

In this section, we identify the optical counterparts of H\,{\sc i}
galaxy groups. To ensure the reliability of the optical counterparts,
we require that all galaxies have optical spectroscopic redshifts. To
ensure the homogeneity of the optical parameters, we restrict the
identification of optical counterparts to the DESI DR1 footprint. The
DESI DR1 covers approximately 9,500\,deg$^2$ and contains about 13.1
million spectra of galaxies \citep{daa+26}. The primary goal of the
DESI survey is to cover galaxies brighter than $m_r = 19.5$ mag
\citep{daa+16}. Note that DESI does not re-observe some bright sources
within its footprint; therefore, we use spectra from SDSS DR18
\citep{aaa+23}, 2MASS \citep{hmm+12}, and NED to supplement these
bright galaxies. To assess the completeness of our combined
spectroscopic sample, we compare it with spectroscopic data from the
GAMA fields \citep{dbr+22}. The comparison shows that the completeness
of our sample is about 90\% at $m_r = 18$ mag and still reaches 83\%
at $m_r = 19.5$ mag \citep{ywh26}. We calculate stellar masses of
galaxies using the empirical relation from \citet{wh24}, based on the
$z$- and $W1$-band luminosities and the $r-z$ color. We compare our
mass estimates with the MPA-JHU results based on SDSS data
\citep[e.g.,][]{khw+03,bcw+04}, and find that the scatter is less than
0.1 dex for galaxies at $z < 0.1$.

Starting with the center of each H\,{\sc i} group, we select optical
galaxies corrosponding to each H\,{\sc i} group from the optical
galaxy catalog, using the maximum projected offset, \(\Delta
r_{\max}\), and the maximum velocity offset, \(\Delta v_{\max}\), of
its H\,{\sc i} members as selection thresholds. Note that DESI may
observe different regions of a certain nearby galaxy, so these
duplicate observations need to be removed. We consider two
spectroscopic observations to be duplicates of the same galaxy if
their projected separation is less than 10\,kpc and their redshift
difference is less than 0.001, and we retain only the observation with
the larger stellar mass. We find that among the 459 H\,{\sc i} groups,
340 systems lie within the DESI DR1 footprint, and 335 of them have
corresponding galaxies in the optical spectroscopic data. We inspected
the photometric images of the remaining 5 groups; all of them exhibit
obvious galaxy clustering, but they are all located near the boundary
of the DESI DR1 footprint, so the absence of spectroscopically
identified optical counterparts may be due to incomplete coverage or
insufficient exposure depth. The number of member galaxies confirmed
by optical spectroscopy, \(N_{\rm s}\), and their total stellar mass,
\(M_{*,\rm tot}\), are listed in Table~\ref{tab1} respectively for
each H\,{\sc i} group. As an example, the optically confirmed member
galaxies of the group with $\rm ID=1$ are marked by small circles in
Figure~\ref{fig1}.

\section{Results}
\label{sect3}
\subsection{Redshift and mass distributions of H\,{\sc i} galaxy groups}

The Left panel of Figure~\ref{fig2} shows that the sample redshifts
are mainly concentrated at \(z<0.025\), with a peak at \(z\sim0.02\)
and a sharp cutoff toward higher redshifts ($z_{\rm
  max}\sim0.05$). This distribution is jointly modulated by the
H\,{\sc i} survey flux limit and the membership threshold of our group
sample. Low-mass H\,{\sc i} sources have insufficient signal-to-noise
ratio at higher redshifts. In addition, because we require H\,{\sc i}
groups to contain at least five members, the insufficient detection of
faint H\,{\sc i} members further limits group identification at higher
redshift. FASHI DR2 \citep{zzj+26}, with the high sensitivity of FAST,
robustly constrains the HIMF down to $M_{\mathrm{HI}}\sim
10^{6.2}M_{\odot}$ over a redshift range of \(z<0.09\); however, it
remains a flux-limited survey, so low-mass galaxies can only be
detected within a smaller comoving volume, while high-mass galaxies
can be identified over a larger volume. Therefore, the redshift peak
at \(z\sim0.02\) in our sample jointly reflects the detection depth
and membership threshold, rather than the intrinsic peak of the
redshift distribution of H\,{\sc i} galaxy groups.

The right panel of Figure~\ref{fig2} shows that the total H\,{\sc i}
mass of galaxy groups in our sample exhibits a nearly symmetric,
single-peaked distribution, with the median peak concentrated at
$M_{\rm{HI,tot}}\approx 10^{10}~M_{\odot}$ and an overall mass range
roughly covering \(10^{9}\) to \(10^{11}~M_{\odot}\). This
distribution is different from the H\,{\sc i} mass function (HIMF) of
individual H\,{\sc i} sources. The global HIMF knee mass measured by
FASHI DR2 is $M_{\rm HI}=10^{9.89\pm0.02}~M_{\odot}$. The systematic
study of the group galaxy HIMF by \citet{jha+20} found that the knee
mass (\(\sim 10^{10}\)) is slightly higher than the global value
obtained by their earlier work \citep{jhg+18}. Our group total mass
distribution does not show a significant low-mass tail, indicating
that the total gas reservoir is still dominated by high-mass galaxies.

\subsection{H\,{\sc i}-stellar mass relation and H\,{\sc i}-stellar mass ratio}

\begin{figure}
  \centering
      {\includegraphics[angle=0,width=0.42\textwidth]{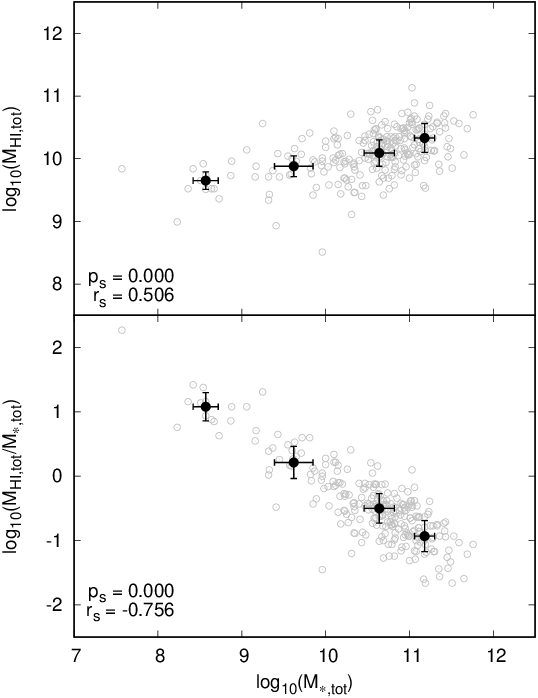}}
      \caption{The H\,{\sc i}--stellar mass relation (upper panel) and
        H\,{\sc i}--stellar mass ratio (lower panel) of H\,{\sc i}
        galaxy groups. Open circles represent individual groups,
        filled circles indicate the medians of each stellar-mass bins,
        and the error bars are calculated through the Median Absolute
        Deviation (MAD) method.}
      \label{fig3}
\end{figure}

Now we investigate how the total H\,{\sc i} mass within galaxy groups
varies with its total mass. We adopt the total stellar mass of
optically confirmed member galaxies as a proxy for the group total
mass. Our spectroscopic sample is $\sim$83\% complete at \(m_{\rm r} <
19.5\) mag, and about 97\% of our identified optical members are
brighter than this limit, indicating that the derived total stellar
mass is only weakly affected by spectroscopic incompleteness and can
be used as a reliable proxy of group total mass. The upper panel of
Figure~\ref{fig3} shows that the total H\,{\sc i} mass of galaxy
groups is positively correlated with the total stellar mass. The
significance of Spearman rank ordered correlation satisfies \(p_s
\approx 0.000\) \citep[see definition in][]{ptv+92}, indicating the
correlation is significant. \citet{gjh+20}, by stacking ALFALFA
groups, found that the total H\,{\sc i} increases with halo mass; the
increase is significant at low richness but becomes gradual when \(N_g
\geq 5\). Our group sample has at least five H\,{\sc i} members,
corresponding to the highest-richness subset in their sample. In this
high-richness regime, we still detect a significant increase of the
total H\,{\sc i} mass with total stellar mass, indicating that this
positive correlation does not disappear in high-richness groups. The
H\,{\sc i}-to-halo mass (HIHM) relation of \citet{ddm+23} based on
GAMA groups shows that the group total H\,{\sc i} mass continues to
increase with halo mass up to \(10^{13.7}M_{\odot}\), which is
consistent with our result.

The lower panel of Figure~\ref{fig3} shows that the total H\,{\sc
  i}-to-total-stellar mass ratio decreases significantly with
increasing total stellar mass (\(p_s \approx 0.000\), \(r_s \approx
-0.756\)). This negative correlation is consistent with the finding of
\citet{ddm+23} that the group H\,{\sc i} mass fraction decreases with
increasing halo mass. \citet{gjh+20} pointed out that satellite
galaxies dominate the H\,{\sc i} content in halos with \(M_{\rm h} >
10^{12.5}h^{-1}M_{\odot}\), and satellites are more susceptible to
ram-pressure stripping and tidal stripping, leading to a decline in
the overall gas fraction. The study of the group galaxy H\,{\sc i}
mass function by \citet{jha+20} also showed that the low-mass slope
for group galaxies is significantly flatter, reflecting environmental
suppression of H\,{\sc i}-poor galaxies. Therefore, our results are
consistent with the picture of gradual H\,{\sc i} depletion in groups:
the total H\,{\sc i} mass increases with group stellar mass, while the
relative gas content continuously declines.

\section{Summary}
\label{sect4}

H\,{\sc i} galaxy groups are key laboratories for understanding galaxy
evolution and environmental effects, and constructing group samples
selected directly from blind H\,{\sc i} surveys is essential for
avoiding optical selection bias and for systematically studying the
gas content in groups. We construct the first blind H\,{\sc
  i}-selected galaxy group sample from the FASHI DR2. Applying a
group-finding algorithm with a density contrast criterion and a
minimum membership of five, we obtain 459 galaxy groups containing
3015 H\,{\sc i} sources. Optical counterparts are identified using
spectroscopic data from DESI, SDSS, 2MASS, and NED, and the total
stellar mass of member galaxies is adopted as a proxy for the total
mass of the host group. The group redshifts are mainly concentrated at
\(z<0.025\), with a peak at \(z\sim0.02\). The total H\,{\sc i} mass
distribution is nearly symmetric and single-peaked, with a peak at
\(\sim10^{10}M_{\odot}\) and a range of roughly
\(10^{9}\)--\(10^{11}M_{\odot}\).

We further investigate the relation between the total H\,{\sc i} mass
and the total stellar mass of groups. The two quantities are
significantly positively correlated, indicating that richer groups
host more H\,{\sc i} gas, consistent with previous stacking
results. Meanwhile, the ratio of total H\,{\sc i} mass to total
stellar mass decreases significantly with increasing total stellar
mass, indicating a gradual decline in the relative H\,{\sc i} content
in group environments. This sample provides a rich target set for
future studies of galaxy evolution and large-scale structure.

\begin{acknowledgments}

Z.S.Y. thanks Prof. Jinlin Han and Dr. Tao Hong for helpful
discussions. This work is supported by the Specialized Research Fund
for State Key Laboratory of Radio Astronomy and Technology, the
National Natural Science Foundation of China (Grant No. 12588202), the
Chinese Academy of Sciences via project JZHKYPT-2021-06, the Science
Research Grants from the China Manned Space Project (Grant
No. CMS-CSST-2025-A04), and the National SKA Program of China (Grant
No. 2022SKA0120103). Z.L.W. is supported by the National Astronomical
Observatories of the Chinese Academy of Sciences (No. E4ZR0506).

This work made use of the data from FAST (Five-hundred-meter Aperture
Spherical radio Telescope). FAST is a Chinese national mega-science
facility, operated by National Astronomical Observatories, Chinese
Academy of Sciences.

This research used data obtained with the Dark Energy Spectroscopic
Instrument (DESI). DESI construction and operations is managed by the
Lawrence Berkeley National Laboratory. This material is based upon
work supported by the U.S. Department of Energy, Office of Science,
Office of High-Energy Physics, under Contract No. DE–AC02–05CH11231,
and by the National Energy Research Scientific Computing Center, a DOE
Office of Science User Facility under the same contract. Additional
support for DESI was provided by the U.S. National Science Foundation
(NSF), Division of Astronomical Sciences under Contract
No. AST-0950945 to the NSF’s National Optical-Infrared Astronomy
Research Laboratory; the Science and Technology Facilities Council of
the United Kingdom; the Gordon and Betty Moore Foundation; the
Heising-Simons Foundation; the French Alternative Energies and Atomic
Energy Commission (CEA); the National Council of Humanities, Science
and Technology of Mexico (CONAHCYT); the Ministry of Science and
Innovation of Spain (MICINN), and by the DESI Member Institutions:
www.desi.lbl.gov/collaborating-institutions. The DESI collaboration is
honored to be permitted to conduct scientific research on I’oligam
Du’ag (Kitt Peak), a mountain with particular significance to the
Tohono O’odham Nation. Any opinions, findings, and conclusions or
recommendations expressed in this material are those of the author(s)
and do not necessarily reflect the views of the U.S. National Science
Foundation, the U.S. Department of Energy, or any of the listed
funding agencies.

This publication makes use of data products from 2MASS, which is a
joint project of the University of Massachusetts and the Infrared
Processing and Analysis Center/California Institute of Technology,
funded by the National Aeronautics and Space Administration and the
National Science Foundation. This research has made use of the
NASA/IPAC Extragalactic Database (NED), which is funded by NASA and
operated by the California Institute of Technology. Funding for the
Sloan Digital Sky Survey IV has been provided by the Alfred P. Sloan
Foundation, the U.S. Department of Energy Office of Science, and the
Participating Institutions. SDSS-IV acknowledges support and resources
from the Center for High-Performance Computing at the University of
Utah. The SDSS website is \url{www.sdss.org}.

\end{acknowledgments}



%


\appendix

The H\,{\sc i} members listed in Table~\ref{tab2} are identified
through the group-finding algorithm described in Section~2.2. Starting
from each H\,{\sc i} source in the FASHI DR2 catalog as a potential
group center, we search for companions within a projected radius of
\(300~\mathrm{kpc}\) and a velocity window of
\(500~\mathrm{km\,s^{-1}}\). To reject spurious associations, we
require that the source density within \(300~\mathrm{kpc}\) be at
least three times that within \(1~\mathrm{Mpc}\). Candidate groups
sharing members are merged, and the group center is defined as the
density center. We then perform three iterations of refinement,
updating the search radius and velocity window based on the dynamical
mass and escape velocity, and applying the density contrast criterion
at each step. Finally, groups with at least five H\,{\sc i} members
are retained. The H\,{\sc i} sources that remain associated with each
group after these iterations constitute the members listed in
Table~\ref{tab2}, and their parameters are taken directly from the
FASHI DR2 catalog \citep{zzj+26}.

\begin{deluxetable*}{crrrrrrrrr}
\tabletypesize{\footnotesize}
\tablecaption{Parameters of H\,{\sc i} members of 459 galaxy groups (the full sample is available at the CDS).}
\tablehead{
\colhead{Group} & \colhead{Source name} & \colhead{RA} & \colhead{DEC} & \colhead{$z_{\rm opt}$} & \colhead{snr} & \colhead{$v_{\rm rad}$} & \colhead{$D$} & \colhead{log$_{10}M_{\rm HI}$} & \colhead{$C$} \\
\colhead{ID} & & \colhead{(J2000)} & \colhead{(J2000)} & & \colhead{($\rm km~s^{-1}$)} & \colhead{(Mpc)} & \colhead{($M_{\odot}$)} & & \\
\colhead{(1)} & \colhead{(2)} & \colhead{(3)} & \colhead{(4)} & \colhead{(5)} & \colhead{(6)} & \colhead{(7)} & \colhead{(8)} & \colhead{(9)} & \colhead{(10)}}
\startdata
 1 & J000113.25$+$130611.9 & 0.30522 & 13.10330 & 0.01782 & 41.3 & 5250.2$\pm$1.4 & 72.1$\pm$5.4 & 10.03$\pm$0.07 & 1.000 \\
 1 & J000104.16$+$130333.8 & 0.26733 & 13.05939 & 0.01789 & 37.5 & 5269.6$\pm$0.8 & 72.5$\pm$5.4 &  9.49$\pm$0.07 & 0.993 \\
 1 & J000145.43$+$130524.8 & 0.43931 & 13.09022 & 0.01836 &  8.9 & 5404.6$\pm$6.1 & 74.5$\pm$5.6 &  8.97$\pm$0.07 & 0.164 \\
 1 & J000127.18$+$125459.3 & 0.36324 & 12.91646 & 0.01859 &  7.4 & 5470.8$\pm$3.4 & 75.4$\pm$5.7 &  8.36$\pm$0.07 & 0.184 \\
 1 & J000140.10$+$125441.0 & 0.41708 & 12.91139 & 0.01754 & 19.3 & 5168.2$\pm$1.0 & 66.6$\pm$5.0 &  8.64$\pm$0.07 & 0.845 \\
 1 & J000220.10$+$125818.0 & 0.58375 & 12.97167 & 0.01840 & 38.1 & 5417.6$\pm$1.4 & 74.7$\pm$5.6 &  9.90$\pm$0.07 & 0.992 \\
 1 & J000046.15$+$124638.7 & 0.19231 & 12.77742 & 0.01778 & 14.8 & 5236.5$\pm$1.3 & 71.6$\pm$5.4 &  8.87$\pm$0.07 & 0.932 \\
 1 & J000251.40$+$130556.0 & 0.71417 & 13.09889 & 0.01742 &  9.2 & 5132.7$\pm$2.5 & 65.2$\pm$4.9 &  9.07$\pm$0.07 & 0.652 \\
 2 & J000127.71$+$312956.7 & 0.36546 & 31.49907 & 0.01611 & 24.3 & 4752.7$\pm$1.5 & 60.6$\pm$4.5 &  8.68$\pm$0.07 & 0.506 \\
 2 & J000137.32$+$313119.1 & 0.40549 & 31.52198 & 0.01577 &  9.9 & 4653.8$\pm$1.8 & 56.3$\pm$4.2 &  7.93$\pm$0.07 & 0.344 \\
 2 & J000112.47$+$313041.5 & 0.30195 & 31.51153 & 0.01656 & 19.7 & 4883.6$\pm$1.8 & 68.0$\pm$5.1 &  8.51$\pm$0.07 & 0.742 \\
 2 & J000129.88$+$312631.9 & 0.37448 & 31.44220 & 0.01603 & 57.4 & 4730.1$\pm$1.0 & 60.1$\pm$4.5 &  9.13$\pm$0.07 & 0.991 \\
 2 & J000145.26$+$312321.2 & 0.43860 & 31.38921 & 0.01589 & 16.4 & 4689.3$\pm$2.5 & 59.4$\pm$4.5 &  8.80$\pm$0.07 & 0.329 \\
 3 & J000448.20$-$013405.3 & 1.20084 & -1.56815 & 0.02390 & 40.8 & 6997.5$\pm$0.8 & 96.8$\pm$4.8 & 10.06$\pm$0.04 & 0.975 \\
 3 & J000447.05$-$012946.1 & 1.19603 & -1.49613 & 0.02393 & 38.2 & 7006.1$\pm$1.3 & 96.9$\pm$4.8 &  9.71$\pm$0.04 & 0.984 \\
 3 & J000455.19$-$014103.7 & 1.22994 & -1.68437 & 0.02392 & 43.1 & 7003.3$\pm$1.2 & 96.9$\pm$4.8 & 10.16$\pm$0.04 & 1.000 \\
 3 & J000436.61$-$014140.5 & 1.15253 & -1.69457 & 0.02383 & 21.0 & 6978.2$\pm$2.9 & 96.6$\pm$4.8 &  9.56$\pm$0.04 & 0.894 \\
 3 & J000418.92$-$012725.5 & 1.07884 & -1.45709 & 0.02427 &  7.8 & 7104.8$\pm$5.1 & 98.2$\pm$4.9 &  8.61$\pm$0.04 & 0.041 \\
 ... & ... & ... & ... & ... & ... & ... & ... & ... & ... \\
 \enddata
\tablecomments{Columns: (1) Group ID. The first row of each group corresponds to the galaxy located at the density center. (2) Source name as defined in the FASHI2 catalog. (3--4) Right ascension and declination of the source; (5) Redshift; (6) Signal-to-noise ratio; (7) Radio velocity; (8) Luminosity distance; (9) H\,{\sc i} mass of the source; (10) Source completeness. The parameters in Columns (2--10) are all taken from the FASHI DR2 catalog; for their meanings and definitions, please refer to \citet{zzj+26}, and additional parameters for these sources can also be obtained from the FASHI DR2 catalog.}
\end{deluxetable*}


\bibliography{ref}{}
\bibliographystyle{aasjournalv7}



\end{document}